\documentclass[a4paper, amsfonts, amssymb, amsmath, reprint, showkeys, footinbib, twoside,superscriptaddress,floatfix,longbibliography]{revtex4-1}

\usepackage{graphicx}%
\usepackage{dcolumn}%
\usepackage{bm}%
\usepackage{float}
\usepackage{xcolor}
\usepackage{mhchem}
\usepackage{gensymb}
\usepackage{ulem}
\usepackage{verbatim}
\usepackage{newtxtext}
\usepackage[slantedGreek]{newtxmath}
\usepackage{amsmath}

\usepackage[T1]{fontenc}
\usepackage{hyphenat}
\usepackage{comment}
\usepackage[pdftex, bookmarks, pdffitwindow=false, pdfstartview=FitH, pdfdisplaydoctitle, colorlinks, plainpages=false, pdftitle={},pdfauthor={}, pdfpagelabels, hypertexnames, citecolor={blue!50!black},linkcolor={blue!50!black}, urlcolor={blue!50!black}, pdflang={en}, hyperfootnotes=false, breaklinks]{hyperref}

\usepackage{siunitx}
\usepackage{comment}

\DeclareUnicodeCharacter{0308}{HERE!HERE!}

\DeclareSIUnit\angstrom{\text {Å}}

\makeatletter
\renewcommand{\@seccntformat}[1]{}
\makeatother

\makeatletter
\def\@bibdataout@aps{
 \immediate\write\@bibdataout{
 @CONTROL{
   apsrev41Control, author="48",editor="1",pages="0",title="0",year="1"
 }}
 \if@filesw
  \immediate\write\@auxout{\string\citation{apsrev41Control}}
 \fi
}
\makeatother

\begin{document}

\preprint{}

\title{Thermal transport of amorphous hafnia across the glass transition}

\author{Zezhu Zeng}
\email{zzeng@ist.ac.at}
\affiliation{The Institute of Science and Technology Austria, Am Campus 1, 3400 Klosterneuburg, Austria}

\author{Xia Liang}
\affiliation{Department of Materials, Imperial College London, South Kensington Campus, London SW7 2AZ, United Kingdom}

\author{Zheyong Fan}
\affiliation{College of Physical Science and Technology, Bohai University, Jinzhou 121013, China}

\author{Yue Chen}
\affiliation{Department of Mechanical Engineering, The University of Hong Kong, Pokfulam Road, Hong Kong SAR, China}

\author{Michele Simoncelli}
\affiliation{Theory of Condensed Matter Group of the Cavendish Laboratory, University of Cambridge, United Kingdom}
\affiliation{Department of Applied Physics and Applied Mathematics, Columbia University, New York (USA)}

\author{Bingqing Cheng}
\email{bingqingcheng@berkeley.edu}
\affiliation{Department of Chemistry, University of California, Berkeley, CA, USA}
\affiliation{The Institute of Science and Technology Austria, Am Campus 1, 3400 Klosterneuburg, Austria}

\date{\today}
\begin{abstract}
Heat transport in glasses across a wide range of temperature is vital for applications in gate dielectrics and heat insulator. 
However, it remains poorly understood due to the challenges of modeling vibrational anharmonicity below glass transition temperature and capturing configurational dynamics across the transition. 
Interestingly, recent calculations predicted that amorphous hafnia (a-HfO$_2$) exhibits an unusual drop in thermal conductivity ($\kappa$) with temperature, contrasting with the typical rise or saturation observed in glasses upon heating. 
Using molecular dynamics simulations with a machine-learning-based neuroevolution potential, we compute the vibrational properties and $\kappa$ of a-HfO$_2$ from 50~K to 2000~K. 
At low temperatures, we employ the Wigner transport equation to incorporate both anharmonicity and Bose-Einstein statistics of atomic vibration in the calculation of $\kappa$. 
At above 1200~K, atomic diffusion breaks down the Lorentzian-shaped quasiparticle picture and makes the lattice-dynamics treatment invalid. 
We thus use molecular dynamics with the Green-Kubo method to capture convective heat transport in a-HfO$_2$ near the glass transition at around 1500~K. 
Additionally, by extending the Wigner transport equation to supercooled liquid states, we find the crucial role of low-frequency modes in facilitating heat convection. 
The computed $\kappa$ of a-HfO$_2$, based on both Green-Kubo and Wigner transport theories, reveals a continuous increase with temperature up to 2000~K.
\end{abstract}

\maketitle



Understanding the thermal conductivity ($\kappa$) of glasses across the glass transition is crucial for applications such as electronic devices \cite{cahill2014nanoscale}, thermal barrier coatings \cite{shelby2020introduction}, and nuclear industries \cite{shelby2020introduction}. 
However, the heat transport mechanisms in glasses, from cryogenic to extremely high temperatures, remain poorly understood;
unlike crystals, glasses lack long-range atomic ordering and the typical phonon picture breaks down~\cite{deangelis2019thermal}. 

Experimentally, 
reported $\kappa$ for glasses are often constrained to thin films \cite{cahill1987thermal, cahill2003nanoscale} as films are easier to vitrify due to faster cooling rates compared to bulk glass in preparation,
and high-temperature thermal measurements \cite{zhao2016measurement} are challenging, limiting data to small systems and narrow temperature ranges. 
For example, measurements of $\kappa$ for glass a-SiO$_2$ \cite{cahill1987thermal, cahill2003nanoscale} span from 50~K to 1200~K, which remains below its glass transition temperature of $\sim$1500~K \cite{polian2002elastic}. 

Computationally, previous studies have mainly used molecular dynamics (MD) with empirical force fields, 
such as Beest-Kramer-Santen \cite{larkin2014thermal} and Tersoff \cite{lv2016non} potentials for a-SiO$_2$, 
Stillinger-Weber \cite{larkin2014thermal, moon2018propagating} and Tersoff \cite{lv2016direct} potentials for a-Si. 
While efficient, these potentials lack quantum mechanical accuracy, resulting in apparent deviations from experiments \cite{zhang2022thermal}.
$ab~initio$ MD (AIMD) \cite{car1985unified} offers higher precision, but it is limited to small systems and short timescales, making it particularly improper for complex amorphous systems. 
Recently, machine learning potential (MLP) \cite{deringer2019machine} has emerged as a promising alternative, delivering density functional theory (DFT)-level accuracy with high computational efficiency. 
For example, Sivaraman et al. \cite{sivaraman2020machine, sivaraman2021experimentally} developed a Gaussian approximation potential (GAP) \cite{bartok2010gaussian} for amorphous HfO$_2$ (a-HfO$_2$), enabling large-scale MD simulations with nearly quantum mechanical precision. 

\begin{figure*}[htb]
    \centering
    \includegraphics[scale=0.78]{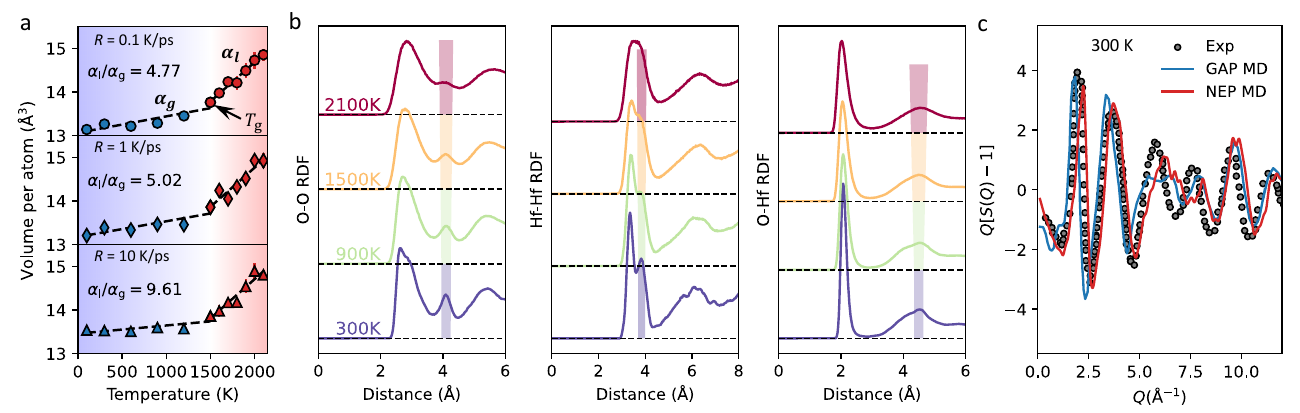}
    \caption{\textbf{a} Temperature dependence of the volume per atom for a-HfO$_2$ prepared at three quench cooling rates ($R$ = 0.1, 1 and 10 K/ps). \textbf{b} Radial distribution functions (RDF) for different atomic pairs (O-O, Hf-Hf, and O-Hf), calculated from NEP MD simulations at temperatures ranging between 300~K and 2100~K.
    \textbf{c} X-ray structure factors computed from NEP MD simulations, and the comparison with previous GAP MD simulations \cite{sivaraman2020machine} and experiment \cite{gallington2017structure}.}
    \label{fig:fig1}
\end{figure*}

Another computational challenge lies in developing accurate theories for computing and understanding $\kappa$ of glasses.
The well-known Allen-Feldman theory \cite{allen1989thermal,allen1993thermal} neglects vibrational anharmonicity. 
Green-Kubo method combined with MD simulations can capture full-order anharmonicity but fails to account for quantum Bose-Einstein statistics. 
Recently, advanced methods within the lattice dynamics (LD) framework, such as the Wigner transport equation (WTE) \cite{simoncelli2019unified,simoncelli2022wigner,caldarelli2022many} and its regularization (rWTE) \cite{simoncelli2023thermal}, as well as quasi-harmonic Green-Kubo (QHGK) theories \cite{isaeva2019modeling,fiorentino2023green} with hydrodynamic extrapolation \cite{fiorentino2023hydrodynamic}, have been developed to compute $\kappa$ in finite-size models of disordered materials accounting for anharmonicity, structural disorder, and Bose–Einstein statistic of vibrations.
These methods open doors for reliably studying glass heat transport, including a-SiO$_2$ \cite{simoncelli2022wigner} (25~K to 1200~K), a-Al$_2$O$_3$ \cite{harper2024vibrational} (50~K to 700~K), and a-Si \cite{isaeva2019modeling} (50~K to 1200~K). 
However, these studies are limited to temperatures below the glass transition.
The exploration of $\kappa$ in glasses across the glass transition remains in its infancy.

In this work, we focus on a-HfO$_2$, renowned for its excellent thermal stability \cite{sivaraman2021experimentally} and applications in resistive random-access memory \cite{broglia2014molecular}. 
Using MLP MD simulations, we computed its dynamical and heat transport behaviors from 50~K to 2000~K, covering its glass transition temperature near 1500~K.
Understanding its thermal and dynamical properties is critical for ensuring the stability and reliability in advanced applications. 
In addition, a previous study \cite{zhang2023vibrational}, based on QHGK theory, reported an decrease of computed $\kappa$ in a-HfO$_2$ with increasing temperature, contrasting the typical trend of $\kappa$ increasing or plateauing in glasses, warranting further investigation.
We first trained a MLP based on the neuroevolution potential (NEP) approach~\cite{fan2021neuroevolution, song2024general}, using the training set from Sivaraman et al. \cite{sivaraman2020machine,sivaraman2021experimentally}.
We prepared the structures of a-HfO$_2$ by melt-quench-anneal process in NEP MD simulations, and then characterized their structure and vibrational properties across its glass transition temperature. 
We then computed its $\kappa$ over a wide range of temperatures, by using Green-Kubo theory, as well as by applying and extending rWTE \cite{simoncelli2023thermal}.

\paragraph{\textbf{Glass transition and atomic diffusion}}

We first prepared glass structures in simulations. 
The melt-quench method in MD is commonly used to generate glass structures \cite{deringer2021origins}, although the simulated cooling rate (1$\sim$10~K/ps) is much higher than experimental rate ($\sim$10~K/s), often yielding higher-energy, non-equilibrium structures. 
Particle swap Monte Carlo methods \cite{ninarello2017models, berthier2023modern} can offer improved equilibration by accelerating phase space sampling with non-local moves, but are limited to the application of polydisperse system due to the extremely low acceptance rates in ionic materials like a-HfO$_2$. 
Therefore, we stick to the melt-quench-anneal method for preparing the initial structures (see Note S1 and S2 of Supplementary Information (SI) for details on the cooling rate $R$ used). 

We then estimated the glass transition temperature $T_{\rm g}$ by calculating volume thermal expansion, as its derivative has a distinct jump near $T_{\rm g}$ in glass-forming materials \cite{debenedetti2001supercooled,lunkenheimer2023thermal}. 
Fig.~\ref{fig:fig1}a shows the volume thermal expansion of a-HfO$_2$ from 100~K to 2100~K at three cooling rates, with a shift in temperature dependence around 1500~K, indicating the glass transition. 
Below 1500~K, thermal expansion shows weaker temperature dependence,
while above this temperature and below the melting point 
($ T_{\rm m} \approx $ 3100~K \cite{sivaraman2021experimentally}), a supercooled liquid forms. 
Recently, Lunkenheimer et al. \cite{lunkenheimer2023thermal} observed that the ratio of the volume thermal expansion coefficient in the supercooled liquid ($\alpha_{\rm l}$) to that in the glass ($\alpha_{\rm g}$) is consistently around 3 across 200 glassy materials with $T_{\rm g}$ values from 200~K to 1100~K. 
Interestingly, our simulations show that this ratio in a-HfO$_2$, $\alpha_{\rm l} / \alpha_{\rm g}$, decreases with lower cooling rates (i.e., closer to real cooling rates). 
At $R$ = 0.1~K/ps, we find $ \alpha_{\rm l} / \alpha_{\rm g} = 4.8 $, approaching the reported ratio of 3 and extending the universal expansion law \cite{lunkenheimer2023thermal} to a-HfO$_2$ with a higher $T_{\rm g} \approx$ 1500~K.

\begin{figure*}[htb]
    \centering
    \includegraphics[scale=0.5]{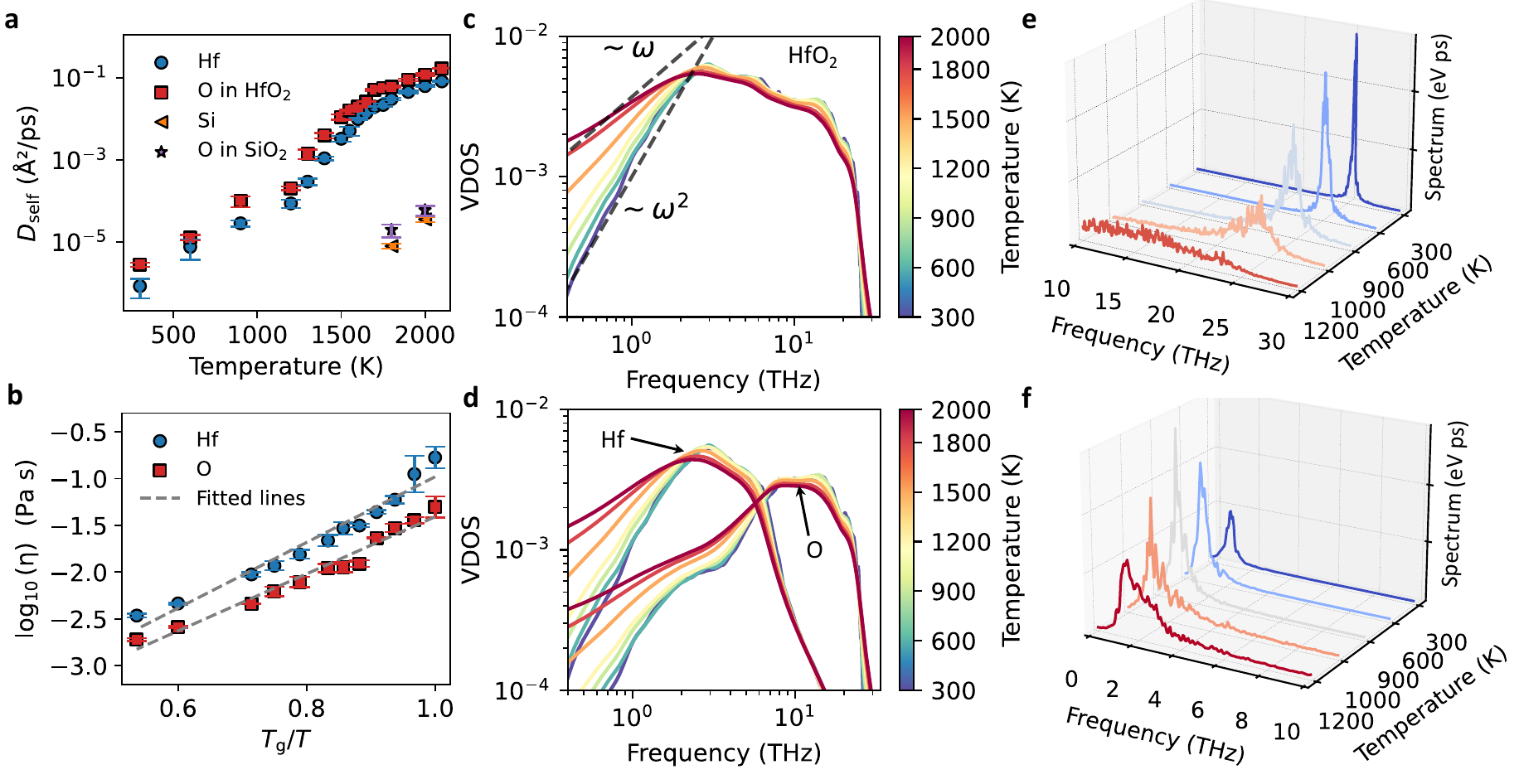}
    \caption{\textbf{a} Atomic self-diffusion coefficients ($D$) of hafnium (Hf) and oxygen (O) in a-HfO$_2$, and silicon (Si) and oxygen in a-SiO$_2$, computed from the velocity autocorrelation function in NEP MD simulations. \textbf{b} Logarithmic viscosity, $\log_{10}(\eta)$, as a function of $T_g/T$ for Hf and O in a-HfO$_2$. Error bars represent uncertainties, with dashed lines indicating linear fits. 
    \textbf{c} Temperature-dependent vibrational density of states (VDOS) of a-HfO$_2$ with respect to frequency ($\omega$) calculated from NEP MD simulations. The dashed lines indicate the $\omega^2$ scaling typical of solids and the $\omega$ scaling characteristic of liquids. 
    \textbf{d} Projected VDOS for Hf and O elements.
    \textbf{e} Temperature-dependent power spectrum of the highest frequency mode.
    \textbf{f} The non-zero lowest frequency mode at the $\Gamma$ point for a-HfO$_2$.}
    \label{fig:fig2}
\end{figure*}

\begin{figure}[htb!]
    \centering
    \includegraphics[scale=0.65]{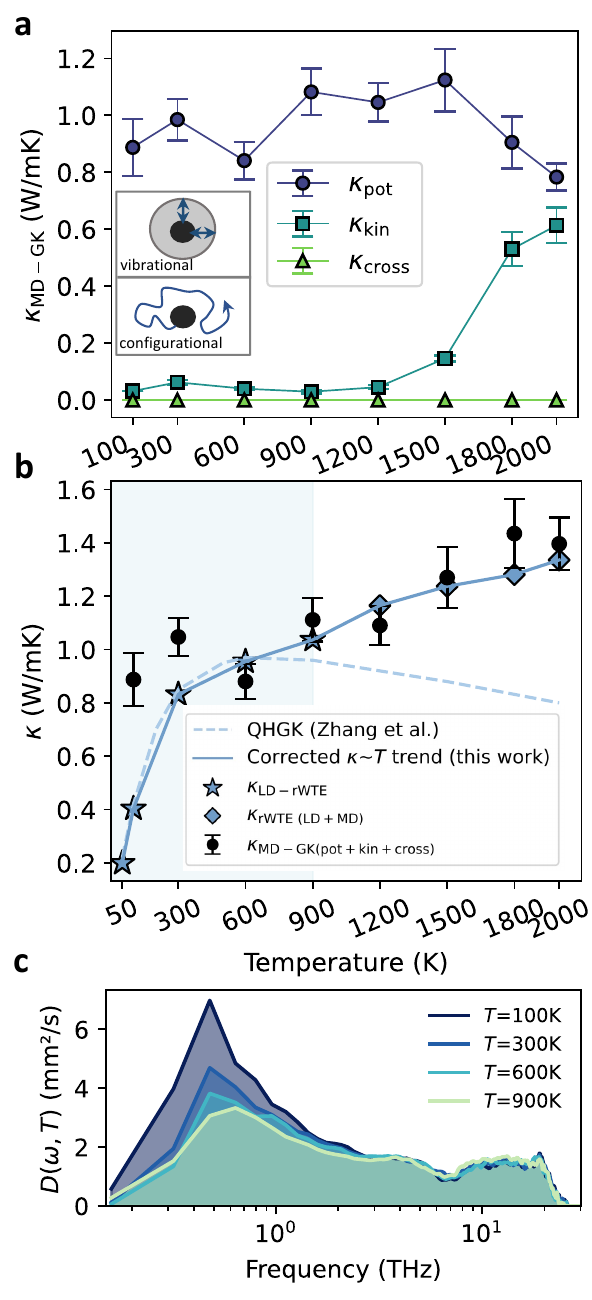}
    \caption{\textbf{a} Temperature dependence of the MD-GK thermal conductivities components for a-HfO$_2$: potential ($\kappa_{\text{pot}}$), kinetic ($\kappa_{\text{kin}}$), and cross ($\kappa_{\text{cross}}$) contributions. The inset panel illustrates the vibrational dynamics (localized vibration indicated by blue double-headed arrows) of a single atom and its configurational dynamics (diffusion indicated by a red solid line). 
    \textbf{b} Comparison of the corrected $\kappa\sim T$ trend (this work) with the QHGK model from Zhang et al. \cite{zhang2023vibrational}, showing $\kappa$ as a function of temperature. The corrected $\kappa$ trend includes results from LD-rWTE ($\kappa_{\text{LD-rWTE}}$) and combined extended rWTE ($\kappa_{\text{rWTE (LD+MD)}}$). 
    The shaded blue region indicates the temperature range where rWTE is applicable (negligible atomic diffusion).
    \textbf{c} Frequency-dependent thermal diffusivity $D(\omega, T)$ computed at temperatures from 100~K to 900~K.
}
    \label{fig:fig3}
\end{figure}

Radial distribution functions (RDFs) illustrated in Fig.\ref{fig:fig1}b) further reveal the structural evolution of a-HfO$_2$.
At low temperatures (e.g., 300~K), the O–Hf RDF features a sharp first peak at $\sim$2~Å, reflecting strong local bonding between O and Hf. 
The Hf–Hf RDF displays two distinct peaks at 3.39~Å and 3.87~Å, corresponding to edge-sharing and corner-sharing polyhedra \cite{sivaraman2020machine,gallington2017structure}, respectively. 
The X-ray structure factor at 300~K (the red curve in Fig.\ref{fig:fig1}c) further highlights typical glass characteristics, showing short-range order and long-range disorder in atomic arrangements. 
The good agreement of structure factors from our NEP MD simulations with experimental data \cite{gallington2017structure} and previous GAP MD simulations \cite{sivaraman2020machine} further confirms the reliability of our NEP model.
As temperature increases to 900~K, these glassy structure features at 300~K remain discernible, although a slight broadening occurs due to increased thermal motion. 
At 1500~K, near the glass transition, the medium-range structural order begins to break down. 
The O–O RDF, which initially shows a distinct second peak (see the trapezoid colorbar in Fig.\ref{fig:fig1}b) indicating medium-range ordering, gradually loses definition and disappears at 2100~K. 
Similarly, the split Hf–Hf peaks merge into a single broad peak at 2100~K. 
This loss of structure, along with the broadening of all RDF features, marks the collapse of medium-range order and the transition to a liquid-like state, reflecting increased atomic diffusion and structural rearrangement.

To understand the dynamics of the glass transition, we examined the self-diffusion coefficients, $D_{\rm self}$, for Hf and O in a-HfO$_2$ (Fig. \ref{fig:fig2}a). 
As temperature rises to 1200~K, $D_{\rm self}$ for both elements increases significantly. 
In a-HfO$_2$, oxygen diffuses slightly faster than hafnium, consistent with previous MD results based on an empirical potential\cite{schie2017ion}. 
Additionally, a-HfO$_2$ exhibits much higher diffusion in the supercooled liquid state than a-SiO$_2$ at 1800~K and 2000~K, suggesting its enhanced atomic mobility. 
Note that the diffusion in supercooled liquid HfO$_2$ remains lower than in liquid HfO$_2$ (0.52 Å$^2$/ps for O and 0.24 Å$^2$/ps for Hf from DFT calculations \cite{hong2018combined}) at 3100~K. 
Fig. \ref{fig:fig2}b shows the viscosity behavior, essential for understanding glass flow and stability near $T_{\rm g}$ \cite{mauro2009viscosity}. The viscosity, calculated via the Stokes-Einstein relation (see SI for details), follows a linear trend with $\log_{10}(\eta)$ versus $T_{\rm g}/T$, suggesting a strong glass behavior with stable structural dynamics across $T_{\rm g}$, in contrast to the abrupt changes seen in fragile glass \cite{mauro2009viscosity}.

To better understand frequency-dependent atomic dynamics, we computed vibrational density of states of a-HfO$_2$ 
(VDOS; see Fig.\ref{fig:fig2}c-d) from velocity autocorrelation functions.
The most pronounced changes occur in the low-frequency (0-3~THz) region, and no distinct temperature difference is observed in the higher-frequency regions. 
At low temperatures (such as 300~K), the VDOS approaches a quadratic scaling relation ($ \sim \omega^2 $) in the low-frequency regime (below 0.8 THz, see the dashed line in Fig.\ref{fig:fig2}c), and displays a mild excess from such Debye scaling (boson peak) around 1 THz. 
As the temperature increases,  the dependence transitions from $\omega^2$ to a linear relationship. Notably, across the $T_{\rm g}$, the VDOS in the low-frequency region becomes linear, a hallmark of liquid-like vibrational behavior \cite{zaccone2021universal}, reflecting a fundamental shift in the dynamics from solid to liquid states. 
The projected VDOS (Fig.\ref{fig:fig2}d) for different elements further elucidates the vibrational contributions: Hf atoms dominate the low-frequency vibrations due to their higher mass and stronger influence on collective acoustic-like modes, 
while O atoms dominate the high-frequency region, associated with localized vibrational modes. 
This elemental partitioning reflects the distinct roles of Hf and O in shaping the vibrational properties and heat transport in a-HfO$_2$.

To further evaluate the impact of atomic diffusion on vibrational mode decay, we computed the power spectrum using the normal mode decomposition method \cite{mcgaughey2014predicting, carreras2017dynaphopy} (see Note S1 of SI).
In Fig.~\ref{fig:fig2}e, we show the power spectrum of the highest-frequency mode in a-HfO$_2$ from 300~K to 1200~K. 
Significant broadening occurs with rising temperature, and at 1200~K, the typical Lorentzian profile is heavily disrupted. As a result, vibrational frequencies and linewidths become ill-defined, indicating that lattice dynamics methods fail to compute $\kappa$ due to the breakdown of the quasiparticle picture. 
Conversely, the lowest-frequency mode (Fig.~\ref{fig:fig2}f) remains well-defined even at 1200~K, where atomic diffusion occurs. 
Low-frequency, phonon-like modes near the Brillouin zone center show greater resilience to breakdown, similar to the robust transverse acoustic modes in superionic conductor AgCrSe$_2$ \cite{ding2020anharmonic} with Ag diffusion.

\paragraph{\textbf{Thermal transport crossover}}

We now discuss $\kappa$ and heat transport mechanisms in a-HfO$_2$.  
Zhang et al. \cite{zhang2023vibrational} trained a NEP model \cite{fan2021neuroevolution, song2024general} using the training set from Sivaraman \cite{sivaraman2020machine} and computed the $\kappa$ of a-HfO$_2$ from 100~K to 2000~K using MD simulations and QHGK approach. 
They reported an unusual trend: 
$\kappa$ continuously decreases with rising temperature above 900~K (see the dashed line in Fig.\ref{fig:fig3}b). 
This behavior contrasts with conventional amorphous materials such as a-Si \cite{zink2006thermal, feldman1993thermal, larkin2014thermal, wang2023quantum, braun2016size} and a-SiO$_2$ \cite{cahill1987thermal, cahill2003nanoscale, lv2016non, zhu2022effect, freeman1986}, 
where both experimental and calculated $\kappa$ increases or saturates with rising temperature. 
Because Zhang et al. \cite{zhang2023vibrational} focus on the heat conduction of a-HfO$_2$ and consider only the three-phonon linewidth of modes in the QHGK method, potential contributions from heat convection and effects of quasiparticle breakdown are not addressed.

We first investigated the critical role of heat convection in a-HfO$_2$ beyond the glass transition.
Fig.\ref{fig:fig3}a presents the temperature-dependent conductivity, $\kappa_{\text{MD-GK}}$, of a-HfO$_2$ from Green-Kubo theory using NEP MD simulations (see Note S2 of SI for more details). 
$\kappa_{\text{MD-GK}}$ is divided into three components: 
$\kappa_{\text{pot}}$ is the conduction via atomic localized vibrations near equilibrium (vibrational dynamics \cite{lunkenheimer2023thermal}, see the inset of Fig.\ref{fig:fig3}a), 
$\kappa_{\text{kin}}$ represents the convection due to atomic diffusion (configurational dynamics \cite{lunkenheimer2023thermal}), 
and $\kappa_{\text{cross}}$ is the cross term between potential and kinetic terms. 
At low temperatures, $\kappa_{\text{pot}}$ dominates, 
and $\kappa_{\text{kin}}$ and $\kappa_{\text{cross}}$ are negligible. 
However, as temperature approaches $T_{\rm g}$, $\kappa_{\text{kin}}$ rises sharply, 
underscoring the role of convection originated from the atomic diffusion in the supercooled liquid state.
In Fig.\ref{fig:fig3}b, we show the $\kappa_{\text{MD-GK}}$ (black circle dots) from 100~K to 2000~K. 
At high temperatures, $\kappa_{\text{MD-GK}}$ increases due to convective contributions, contrasting with the QHGK $\kappa$ reported by Zhang et al. \cite{zhang2023vibrational}. 
As temperature approaches zero, $\kappa_{\text{MD-GK}}$ remains constant, in contrast to the decreasing trend observed in QHGK results. This reflects limitation of classical MD in lacking nuclear quantum effects, leading to constant heat capacity instead of the quantum-reduced values in real materials.

To address this limitation, we employed the rWTE \cite{simoncelli2023thermal} to calculate $\kappa$ (see Note S2 of SI for details) from 50~K to 900~K, as it can intrinsically account for the quantum (Bose-Einstein) distribution of atomic vibrations.
The rWTE method accelerates the convergence of $\kappa$ calculations in strongly disordered solids, where disorder-induced repulsion between vibrational energy levels \cite{simkin2000minimum} is comparable to the intrinsic linewidth caused by anharmonicity or isotopes. 
In such systems, linewidths primarily enable vibrational mode coupling, with minimal impact on the computed $\kappa$. Examples include a-SiO$_2$ \cite{simoncelli2023thermal} and a-Al$_2$O$_3$ \cite{harper2024vibrational}.
The rWTE can be written as \cite{simoncelli2023thermal}:
\begin{equation}
    \label{eq:rwte}
    \begin{aligned}
        \kappa = & \frac{1}{\mathcal{V} N_c} \sum_{\mathbf{q}, s, s^{\prime}} \frac{\omega(\mathbf{q})_s+\omega(\mathbf{q})_{s^{\prime}}}{4}\left(\frac{C(\mathbf{q})_s}{\omega(\mathbf{q})_s}+\frac{C(\mathbf{q})_{s^{\prime}}}{\omega(\mathbf{q})_{s^{\prime}}}\right) \frac{\left\|\mathbf{v}(\mathbf{q})_{s, s^{\prime}}\right\|^2}{3} \\
        & \times \pi \mathcal{F}_{\left[ \Gamma(\mathbf{q})_s + \Gamma(\mathbf{q})_{s'}, \eta \right]} \left( \omega(\mathbf{q})_s - \omega(\mathbf{q})_{s'} \right),
    \end{aligned}
\end{equation}
where $\mathcal{V}$ is the simulation cell volume, $N_c$ the number of $\mathbf{q}$-points, $s$ and $s^{\prime}$ are band indices, $\omega$ the vibrational frequency, $C(\mathbf{q})s$ the quantum-based specific heat, $\mathbf{v}(\mathbf{q}){s, s^{\prime}}$ the velocity operator, and $\Gamma$ the linewidth. The Voigt distribution $\mathcal{F}_{\left[\Gamma(\mathbf{q})s + \Gamma(\mathbf{q}){s^{\prime}}, \eta\right]}$ results from convolving a Lorentzian (width $\Gamma(\mathbf{q})s + \Gamma(\mathbf{q}){s^{\prime}}$) with a Gaussian (variance $\eta^2 \pi / 2$).
Computational details are provided in SI. 

As shown in Fig.\ref{fig:fig3}b, the rWTE $\kappa$ increases with temperature from 50~K to 900~K, consistent with experimental observations for many glasses and also align well with the QHGK results \cite{zhang2023vibrational}. It also agrees with our bare WTE $\kappa$, calculated using a 6144-atom unit cell at $\mathbf{q=0}$ (see Fig. S10 of SI), demonstrating that rWTE can reliably evaluate the heat conduction in a-HfO$_2$ with strong atomic disorder \cite{iwanowski2024bond}. 
Above 900~K, significant atomic diffusion causes a breakdown of the quasiparticle picture (Fig.\ref{fig:fig2}e), rendering LD-based methods like rWTE and QHGK invalid as they both require well-defined frequencies and linewidths for vibrations. Thus, these methods are applicable only in solids where atomic diffusion is negligible and quasiparticle vibrational excitations well defined. 

To shed light on the microscopic mechanisms underlying convective transport at higher temperatures, we phenomenologically employed the WTE framework to supercooled liquid a-HfO$_2$. 
Under the hypothesis that the conductivity of a supercooled liquid can be described by a mathematical expression analogous to that emerging from the WTE, we rewrite Eq.\ref{eq:rwte} into the equivalent form
$\kappa = \frac{1}{\mathcal{V} N_c} \sum_{\mathbf{q}, s} C(\mathbf{q})_s D(\mathbf{q})_s$,
where $ D(\mathbf{q})_s $ represents the anharmonic thermal diffusivity, defined as \cite{simoncelli2023thermal}:
\begin{equation}
\begin{aligned}
D(\mathbf{q})_s = & \sum_{s'} \frac{\omega(\mathbf{q})_s + \omega(\mathbf{q})_{s'}}{2 \left[ \Gamma(\mathbf{q})_s + \Gamma(\mathbf{q})_{s'} \right]} \left[ \frac{C(\mathbf{q})_s}{\omega(\mathbf{q})_s} + \frac{C(\mathbf{q})_{s'}}{\omega(\mathbf{q})_{s'}} \right] \frac{\left\| \mathbf{v}(\mathbf{q})_{s s'} \right\|^2}{3} \\
& \times \pi \mathcal{F}_{\left[ \Gamma(\mathbf{q})_s + \Gamma(\mathbf{q})_{s'}, \eta \right]} \left( \omega(\mathbf{q})_s - \omega(\mathbf{q})_{s'} \right).
\end{aligned}
\end{equation}
To directly compare LD and MD properties, it is informative to express diffusivity as a function of frequency.
Thus, we recast the $D(\mathbf{q})_s$ as a frequency-dependent function as
$
D(\omega, T) = \left[ g(\omega) \mathcal{V} N_c \right]^{-1} \sum_{\mathbf{q}, s} D(\mathbf{q})_s \delta \left( \omega - \omega(\mathbf{q})_s \right)
$,
where $g(\omega)$ represents the VDOS (Fig.\ref{fig:fig2}c).
The frequency-dependent quantities defined above allow to rewrite the rWTE $\kappa$ in frequency space as:
\begin{equation}
    \label{eq:freqkappa}
    \kappa(T) = \int_0^{\infty} g(\omega) C(\omega, T) D(\omega, T) \, d\omega.
\end{equation}

In classical MD simulations at high temperatures, $C(\omega, T)$ becomes temperature-independent and approaches the Boltzmann constant $k_{\text{B}}$. 
Practically, $D(\omega, T)$ also remains nearly temperature-independent in a-HfO$_2$ at these temperatures.
This may be attributed to the interplay between the low-frequency VDOS and the velocity operator, where an increase in one (Fig. \ref{fig:fig2}c) is offset by a decrease in the other, resulting in nearly temperature-independent diffusivity.
Our calculations from 100~K to 900~K (Fig. \ref{fig:fig3}c) reveal minimal differences for $D(\omega, T)$ above 600~K, suggesting that the increase in $\kappa$ in supercooled liquid a-HfO$_2$ likely results from the enhanced low-frequency VDOS. 
Additionally, due to $D(\omega, T)$ has a peak at the low-frequency region, which gives rise to a potentially critical contribution related to low-frequency modes.
We confirm this by integrating the thermal diffusivity from lattice dynamics at 900~K with the temperature-dependent VDOS from MD simulations (Fig. \ref{fig:fig2}c) at higher temperatures using Eq.\ref{eq:freqkappa}. 
As shown in Fig.\ref{fig:fig3}b, the integrated $\kappa$ ($\kappa_{\text{LD+MD}}$) successfully reproduces the 
$\kappa_{\text{MD-GK}}$ results above 900~K. 
This agreement improves our understanding of heat transport across the glass transition in amorphous materials, where the enhanced low-frequency VDOS (dominated by relatively heavy atoms, such as Hf in a-HfO$_2$) plays a key role in increasing $\kappa$. 
From the Green-Kubo perspective within MD, this enhancement arises from heat convection, while the rWTE within LD framework highlights the crucial role of low-frequency propagating modes in enhancing convection in supercooled liquid a-HfO$_2$.

\paragraph{\textbf{Conclusion}}

Through MLP MD simulations, we unveiled the glass transition and structural evolution in a-HfO$_2$, where significant atomic diffusion across the glass transition emerges as a crucial factor shaping thermal transport. 
Using Green-Kubo theory, we quantified the considerable heat convection in a-HfO$_2$.
We further extended the rWTE method to offer new insights into microscopic heat transport mechanisms within the supercooled liquid state. 
Our findings reveal a continuous increase in $\kappa$ of a-HfO$_2$ with temperature up to 2000~K. 
This work deepens understanding of structural transitions and heat transport in amorphous materials, 
and may pave the way for future research in superionic \cite{ding2020anharmonic}, disordered \cite{zeng2024lattice} and supercooled systems.

\textbf{Acknowledgments}

We thank Ludovic Berthier for fruitful discussions and Ting Liang for providing the initial structures of a-SiO$_2$. 
ZZ acknowledges funding from the European Union’s Horizon 2020 research and innovation programme under the Marie Skłodowska-Curie grant agreement No. 101034413. 
The authors also acknowledge the research computing facilities provided by HPC ISTA and ITS HKU.

\nocite{*}
\bibliography{main}
\end{document}